\documentclass[prb,twocolumn,amsmath,amssymb]{revtex4}
\newcommand{\be}{\begin{equation}}
\newcommand{\ee}{\end{equation}}
\newcommand{\ba}{\begin{eqnarray}}
\newcommand{\ea}{\end{eqnarray}}
\newcommand{\nn}{\nonumber \\}

\usepackage{graphicx}
\usepackage{bm}

\begin{document}

\title{Quantum Nondemolition Charge Measurement of a Josephson Qubit}
\author{M.~H.~S.~Amin}
\affiliation{D-Wave Systems Inc., 320-1985 W. Broadway, Vancouver,
B.C., V6J 4Y3 Canada}

\begin{abstract}
In a qubit system, the measurement operator does not necessarily
commute with the qubit Hamiltonian, so that the readout process
demolishes (mixes) the qubit energy eigenstates. The readout time
is therefore limited by such a mixing time and its fidelity will
be reduced. A quantum nondemolition readout scheme is proposed in
which the charge of a flux qubit is measured. The measurement
operator is shown to commute with the qubit Hamiltonian in the
reduced  two-level Hilbert space, even though the Hamiltonian
contains non-commuting charge and flux terms.
\end{abstract}



\maketitle

Reading out the final state of qubits is an essential part of any
quantum computation scheme. Most commonly, the readout is
considered to be projective, meaning that the state of the qubit
is projected onto the eigenstates of the measurement operator.
This, however, requires strong qubit-detector coupling and fast
readout response, which could be practically difficult.
Furthermore, strong interaction with the detector may limit the
coherence time of the system during quantum operation. For a
projective measurement, the dephasing time $\tau_\varphi$ due to
the readout backaction should be much smaller than the free
evolution time of the qubit\cite{shnirman}($\hbar=1$):
$\tau_\varphi \ll \Delta^{-1}$, where $\Delta$ is the qubit's
energy splitting. In general, however, it is also possible to
measure the qubit's state with a weaker coupling \cite{note}
($\tau_\varphi \gg \Delta^{-1}$) as long as $\tau_\varphi \leq
\tau_{\rm meas} \ll \tau_{\rm mix}$. The measurement time
$\tau_{\rm meas}$ determines when the two states have
distinguishable output, and the mixing time $\tau_{\rm mix}$ shows
when the information about the initial state is destroyed due to
the interaction with the detector. In this regime (called
``Hamiltonian dominated'' in Ref.~\onlinecite{shnirman}), the
qubit is measured in the eigenstates of its own Hamiltonian and
the detector's output is the {\em expectation value} of the
measurement operator in these eigenstates. The backaction noise
causes extra relaxation of the qubit to the ground state, limiting
$\tau_{\rm mix}$. The reported single shot readout of flux qubits
using DC-SQUID in Ref.~\onlinecite{takayanagi}, seems to agree
more with this picture than with the projective measurement.

In general, it is desirable to have a long $\tau_{\rm mix}$ to
allow a long measurement time with high readout fidelity. In a
Quantum Nondemolition (QND) measurement, while the readout process
collapses the state of the qubit, it does not demolish
it.\cite{QNDreview} In other words, the readout measures the
eigenstates of the qubit Hamiltonian without changing them
(e.g.~causing fast relaxation). Mathematically, it means that the
measurement operator commutes with the qubit Hamiltonian. Hence,
$\tau_{\rm mix}$ will be limited only by the intrinsic relaxation
time of the qubit, which could be long. While increasing the
readout fidelity, QND measurement allows reducing the
qubit-readout coupling and therefore decreasing unwanted
decoherence due to interaction with the detector during the
operation time.

Despite the importance of QND measurement, there are only few QND
readout schemes for superconducting qubits.\cite{QNDSchemes} In
the Saclay's charge-flux qubit design (quantronium),\cite{vion}
the measurement operator (flux) commutes with the qubit
Hamiltonian only at the flux degeneracy point. At the time of
readout, however, the qubit is moved away from this point, hence
resulting in a readout that is not a QND type. This may be
responsible for the observed small readout fidelity. Another type
of charge-flux qubit, which is flux based, was proposed in
Ref.~\onlinecite{CPQ}. Like in the quantronium case, the qubit can
be optimally protected against dephasing due to low frequency
noises at the charge-flux double degeneracy (so called {\em
magic}) point. In this paper, we suggest a modification to the
qubit, to make QND readout possible.

\begin{figure}[t]
\includegraphics[width=6cm]{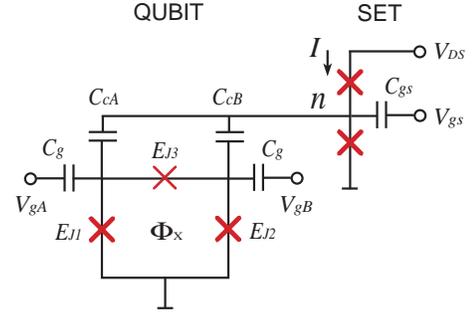}
\caption{Three-Josephson-junction charge-flux qubit coupled to an
SET for QND readout. }\label{fig1}
\end{figure}

The qubit (Fig.~\ref{fig1}) consists of a superconducting loop
containing three Josephson junctions,\cite{mooij} threaded by an
external flux close to half a flux quantum ($\Phi_x \approx
\Phi_0/2 = h/4e$). The Josephson energy and junction capacitance
of two of the junctions are the same ($E_{J1}=E_{J2}=E_J$,
$C_1=C_2=C$) while those of the third junction are slightly
smaller ($\alpha E_J$ and $\alpha C$, with $0.5<\alpha < 1$). In
addition, two voltage sources $V_{gA}$ and $V_{gB}$ are
capacitively connected to two of the islands, which are also
coupled capacitively to an SET or RF-SET \cite{rfSET} as a
sensitive charge detector for readout.\cite{Cqubits} At an
appropriate biasing point, state dependent voltages appear on
these islands,\cite{CPQ} which induce a charge on the SET's
island, affecting the current through it.

We define $\phi=(\phi_1+\phi_2)/2$ and $\theta=(\phi_1-\phi_2)/2$,
where $\phi_{1,2}$ are the phase differences across the two larger
junctions. The Hamiltonian of the system is $H=H_{\rm Q} + H_{\rm
SET}+H_{\rm int}$, where $H_{\rm SET}$ is the SET Hamiltonian, and
\ba
 H_{\rm Q} = {(P_\phi + n_+)^2 \over 2 M_\phi} + {(P_\theta + n_-)^2 \over
 2 M_\theta} + U(\phi,\theta), \nonumber
\ea
is the qubit Hamiltonian. Here
\ba
 U(\phi,\theta) &=& E_J [- 2\cos \phi \cos \theta + \alpha
 \cos (2\pi f + 2\theta)] \nonumber
\ea
is the potential energy, $P_\phi=-i\partial/\partial_\phi$ and
$P_\theta=-i\partial/\partial_\theta$ are the momenta conjugate to
$\phi$ and $\theta$, $M_\phi = C(1+\gamma)/2e^2$, $M_\theta =
C(1+\gamma+2\alpha)/2e^2$, $\gamma=C_g/C$, and $n_\pm = n_{gA} \pm
n_{gB}$, with $n_{gA,gB}=V_{gA,gB}C_g/2e$ being the normalized
gate charges. Throughout this paper we take $f\equiv
\Phi_x/\Phi_0-1/2 =0$, leading to a $U(\phi,\theta)$ with
degenerate minima at $\phi=0$, $\theta=\pm \arccos (1/2\alpha)$.

The interaction Hamiltonian is written as
\ba
 H_{\rm int} &=& \left( {\kappa_+ \over M_\phi} P_\phi
 + {\kappa_- \over M_\theta} P_\theta \right) \hat n.
 \label{deltaH}
\ea
where $\kappa_\pm = (C_{cA}\pm C_{cB})/C_\Sigma$, $\hat n$
represents the normalized charge (operator) of the SET island, and
$C_\Sigma$ is its effective capacitance.

The commutation relation
\ba
 [H_{\rm int}, H_{\rm Q}] &=& -i(\sin \phi \cos \theta)
 {\kappa_+ \over M_\phi} \hat n \nn
 && -i(\cos \phi \sin \theta - 2\alpha \sin 2\theta)
 {\kappa_- \over M_\theta} \hat n  \nonumber \ea
is clearly nonzero, meaning that in general the readout can change
the eigenstates of the qubit Hamiltonian. For a qubit well in the
flux regime, however, the qubit wave function is confined within a
unit cell in the phase space. The fluctuations of $\phi$ is
therefore small and, for the lowest energy eigenstates, $\phi
\approx 0$. Thus, $[H_{\rm int}, H_{\rm Q}] \propto \kappa_-$
vanishes if $C_{gA}=C_{gB}$, suggesting a scheme for QND
measurement. However, inter-unit-cell tunnelling, which is
important for operation in the charge-flux regime,\cite{CPQ} will
delocalize $\phi$, conflicting with the above picture. Therefore,
it is not possible to have $H_{\rm int}$ exactly commute with
$H_Q$. As we shall see however, it is possible to make them
commute in the Hilbert space reduced to the first two energy
levels, hence satisfying a much weaker requirement: $\langle 1 |
H_{\rm int} | 0\rangle = 0$, yet sufficient for QND measurement.
The backaction noise in that case will not affect the rate of
relaxation between the two lowest energy states ($|0\rangle$ and
$|1\rangle$) of the qubit during readout, i.e. will not reduce
$\tau_{\rm mix}$.

To study in more detail, we calculate the relaxation rate due to
the SET's backaction on the qubit:
\ba
 \tau_{\rm mix}^{-1} = \left|{\kappa_+ \over M_\phi}P^{10}_\phi
 + {\kappa_- \over M_\theta}P^{10}_\theta \right|^2
 [S(\Delta)+S(-\Delta)],
 \label{GammaR1}
\ea
where $P^{ij}_{\phi,\theta} \equiv \langle i |P_{\phi,\theta} |
j\rangle$, and
\be
 S(\omega)=\int e^{i\omega t} \langle \hat n(t) \hat n(0) \rangle dt
\ee
is the noise spectral density.  Using the orthodox theory for a
normal SET,\cite{orthodox} the spectral density $S(\omega)$ at
small bias current is given by \cite{korotkov}
%
%
\be
 S(\omega)= {2I/e \over \omega^2 +
 (\Gamma_1 + \Gamma_2)^2}, \label{S}
\ee
where $I=e\Gamma_1 \Gamma_2/(\Gamma_1 + \Gamma_2)$ is the average
current through the SET and $\Gamma_1$ ($\Gamma_2$) is the rate of
tunnelling of an electron into (out of) the island via the first
(second) SET junction:
\ba
 \Gamma_{1,2} &=& {1\over e^2R}{\Delta E_{1,2}/h \over 1-e^{-\Delta
 E_{1,2}/k_BT}}, \\
 \Delta E_{1,2} &=& \pm 2 E_{C_\Sigma}(n_{gs}-1/2)+(1/2)eV_{DS}.
\ea
Here $R$ is the normal resistance of the junctions,
$E_{C_\Sigma}=e^2/2C_\Sigma$ is the charging energy,
$n_{gs}=C_{gs}V_{gs}/e$ is the normalized gate charge, and
$V_{DS}$ is the voltage between drain and source of the SET.

\begin{figure}[t]
\includegraphics[width=8.8cm]{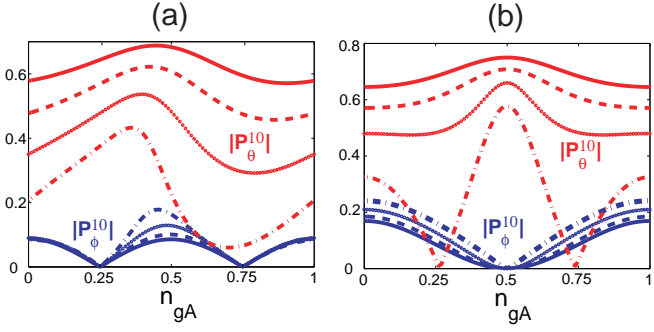}
\caption{$|P_\phi^{10}|$ and $|P_\theta^{10}|$, plotted vs
$n_{gA}$ when $n_{gB}=0.25$ (a) and $n_{gB}=0.5$ (b), for a qubit
with $E_c/E_J=0.2$ at $f=0$. The top curves are for
$|P_\theta^{10}|$ and the bottom ones show $|P_\phi^{10}|$. The
values of $\alpha$ are 0.6 (solid line), 0.7 (deshed line), 0.8
(dotted line), and 0.9 (dash-dot line).}\label{fig2}
\end{figure}

Let us first consider the case where the readout is coupled to
only one of the islands (as in Ref.~\onlinecite{CPQ}): $C_{cB}=0$,
therefore $\kappa_+=\kappa_-=C_{cA}/C_\Sigma$. An SET, operating
in the quasiparticle branch, typically has $I\sim 1$~nA and
$V_{DS}\sim 400~\mu$V, which lead to $\Gamma_{1,2}\sim 10^{10}$
s$^{-1}$. For a typical qubit with $E_C/E_J=0.1$ and
$\alpha=0.75$, as suggested in Ref.~\onlinecite{CPQ}, we find:
$E_C \approx 5$ GHz, $\Delta \approx 7$ GHz, $M_\phi^{-1}\approx
2.5M_\theta^{-1}\approx 4E_C \approx 20$ GHz, $P_\phi^{10}\approx
0.04$, and $P_\theta^{10}\approx 0.5$. Substituting into (\ref{S})
and (\ref{GammaR1}) and using $C_{cA}/C_\Sigma \approx 0.2$, we
find a mixing time of $\tau_{\rm mix} \sim 20$ ns, which is
probably too short to read out the qubit. It is therefore
important to use symmetric coupling (see below) or operate the SET
at the double Josephson quasiparticle point,\cite{DJQP} where the
SET's backaction noise (but also its sensitivity) is significantly
smaller.\cite{Cqubits} Notice that $P_\theta^{10} \gg P_\phi^{10}$
(see also Fig.~\ref{fig2}), hence the second term in
(\ref{deltaH}) has much larger contribution to the backaction
noise than the first one.

In the symmetric coupling case ($C_{cA}=C_{cB}$),  $\kappa_-=0$
and therefore the second term in (\ref{deltaH}) is cancelled. The
relaxation rate will then only depend on $P^{10}_\phi$, which is
typically much smaller than  $P^{10}_\theta$. As we shall see,
there even exist special points at which $P^{10}_\phi=0$, hence
(\ref{GammaR1}) vanishes, making QND readout possible. Let us
write
\be
 P^{10}_\phi = {1\over2} \int d\phi \int d\theta
 \Psi_1(\phi,\theta) (-i\partial_\phi) \Psi_0(\phi,\theta), \label{P10}
\ee
where $\Psi_{0,1}$ are the first two energy eigenfunctions. When
$n_{gA} = n_{gB} = 0$, the wave functions have the following
symmetries
\ba
  \Psi_{0,1}(-\phi,\theta) &=& \Psi_{0,1}(\phi,\theta)
  \label{phisym} \\
  \Psi_{0,1}(\phi,-\theta) &=& \pm \Psi_{0,1}(\phi,\theta) \label{thetasym}
\ea
Either of these is enough to make (\ref{P10}) vanish. The output
however will be zero at this ``magic'' point. It is therefore
necessary to move the qubit away form this point at the time of
readout. As we shall see, unlike in the quatronium case, it is
possible to find points at which (\ref{P10}) vanishes while the
measurement output is finite.

To see explicitly how the symmetries (\ref{phisym}) and
(\ref{thetasym}) change at finite $n_{gA}$ and $n_{gB}$, we first
consider the simpler problem of a one dimensional particle in a
symmetric periodic potential $V(x)$
\ba
 && H= {1\over 2m}(P+eA/c)^2 + V(x), \label{H1d} \\
 && V(x+d) = V(x) = V(-x),
\ea
where $A$ is a uniform vector potential. The eigenfunctions of
this system are described by the Bloch wave functions. The ground
state of the system has zero crystal momentum and is therefore the
solution of (\ref{H1d}) with periodic boundary condition
$\Psi_0(x+d) = \Psi_0(x)$. When $A=0$, the ground state is
symmetric: $\Psi_0(-x) = \Psi_0(x)$. At finite $A$, one can
perform a gauge transformation to remove $A$ from (\ref{H1d}). The
boundary condition then becomes $\Psi_0(x+d) = e^{i\chi}
\Psi_0(x)$, where $\chi = \int (e/\hbar c)A dx=eAd/\hbar c$. When
the wave function is confined within the wells, a standard tight
binding type of calculation can be used to show that a small
antisymmetric term, proportional to $\sin \chi$, will be added to
the otherwise symmetric ground state. In general, the wave
function of the system in the $i$-th state, has the form
\ba
 \Psi_i(x) = \Psi^0_i(x) + \Psi^1_i(x) \sin \chi, \label{sym1d}
\ea
where $\Psi^1_i(x)$ is a small term with opposite symmetry as that
of the main term $\Psi^0_i(x)$. $\Psi^1_i(x)$ is dominantly
determined by the geometry of the well and therefore is
independent of $\chi$. For example, when the two lowest levels of
the system are well separated from the rest, $\Psi^1_0(x) \propto
\Psi^0_1(x)$.

Let us now return to our qubit system. The two dimensional
potential energy of the system is $2\pi$ periodic in both $\phi_1$
and $ \phi_2$ directions. In terms of $\phi$ and $\theta$, the
boundary condition reads
\be
 \Psi(\phi+\pi,\theta \pm \pi) = e^{i\pi(n_+ \pm n_-)}
 \Psi(\phi,\theta).
\ee
A straightforward generalization of the above argument yields
\ba
 && \Psi_i(\phi,\theta) =  \Psi^0_i(\phi,\theta) +
 \Psi^1_i(\phi,\theta) \sin \pi n_+ \nn && \qquad
 + \Psi^2_i(\phi,\theta) \sin \pi n_-
 + \Psi^3_i(\phi,\theta) \sin \pi n_+ \sin \pi n_-. \nonumber
\ea
For the ground (first excited) state [$i=0$ (1)], $\Psi^0_i$ and
$\Psi^2_i$ are symmetric (symmetric) functions of $\phi$, while
$\Psi^1_i$ and $\Psi^3_i$ are antisymmetric (antisymmetric).
Likewise, $\Psi^0_i$ and $\Psi^1_i$ are symmetric (antisymmetric)
in $\theta$, while $\Psi^2_i$ and $\Psi^3_i$ are antisymmetric
(symmetric). The difference between $\phi$ and $\theta$ stems from
the fact that the potential has double-well structure only in
$\theta$ direction. Neglecting the small contribution from the
decay region, equation (\ref{P10}) gives
\ba
 P^{10}_\phi &=& Z \sin \pi n_+ \sin \pi n_-
 \nn &=& {1\over 2} Z (\cos 2\pi n_{gA} - \cos 2\pi n_{gB}), \label{Pphi}
\ea
where $Z = A^{10}_{03} + A^{10}_{12} + A^{10}_{21} + A^{10}_{30}$,
and
\ba
 A^{ij}_{\alpha \beta} = {1\over2} \int d\phi \int d\theta \Psi_i^\alpha (\phi,\theta)
 (-i\partial_\phi) \Psi_j^\beta (\phi,\theta). \nonumber
\ea
Similarly, for $P_\theta$ we find
\ba
 P^{10}_\theta &=& B^{10}_{00} + B^{10}_{11} \sin^2 \pi n_+
 + B^{10}_{22} \sin^2 \pi n_- \nn
 && + B^{10}_{33} \sin^2 \pi n_+ \sin^2 \pi n_-, \label{Ptheta}
\ea
where
\ba
 B^{ij}_{\alpha \beta} = {1\over2} \int d\phi \int d\theta \Psi_i^\alpha (\phi,\theta)
 (-i\partial_\theta) \Psi_j^\beta (\phi,\theta). \nonumber
\ea
While $P^{10}_\theta$ is always finite, $P^{10}_\phi$ vanishes
when either $n_+$ or $n_-$ is an integer number. At these special
points, $\langle 1 | H_{\rm int} | 0\rangle=0$ if $\kappa_-=0$ in
(\ref{deltaH}). The latter can be easily achieved by symmetrically
coupling the two islands of the qubit to the readout SET
($C_{gA}=C_{gB}$).

The above argument is valid beyond the approximations used to
derive (\ref{Pphi}) and (\ref{Ptheta}) (see Fig.~\ref{fig2}).
Indeed, symmetries (\ref{phisym}) and (\ref{thetasym}) always hold
when respectively $n_+$ and $n_-$ are integer numbers, causing
(\ref{P10}) to vanish. A simple way to access these symmetry
points is to adjust the gate voltages so that $n_{gA}= \pm
n_{gB}$, which correspond to $n_\mp = 0$, respectively.

Figure \ref{fig2} shows the result of numerical calculation of
$|P_{\phi,\theta}^{10}|$ for different values of $\alpha$. When
$\alpha$ is small, the wave function of the qubit is localized
within a unit cell. As a result, the $|P_{\phi}^{10}|$ curve
agrees very well with (\ref{Pphi}). A larger $\alpha$ delocalizes
the states, causing deviation from a simple sinusoidal behavior.
When $n_{gB}=0.25$ (Fig.~\ref{fig2}a), $|P_{\phi}^{10}|$ vanishes
at $n_{gA}=0.25$ ($n_-=0$) and $n_{gA}=0.75$ ($n_+=1$), in
agreement with the above argument. For $n_{gB}=0.5$
(Fig.~\ref{fig2}b) both points coincide at $n_{gA}=0.5$ ($n_-=0$
and $n_+=1$).
\begin{figure}[t]
\includegraphics[width=6cm]{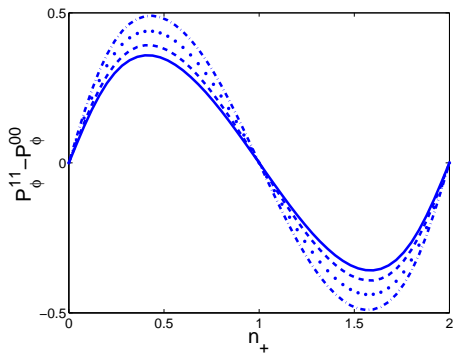}
\caption{$P_\phi^{11}-P_\phi^{00}$ vs $n_+$ for the qubit system
of Fig.~\ref{fig2} when $n_-=0$. Different lines correspond to
different $\alpha$ the way described in the caption of
Fig.~\ref{fig2}. }\label{fig3}
\end{figure}

While the points $n_{gA}=n_{gB}$ and $n_{gA}=-n_{gB}$ both result
in $P^{10}_\phi=0$, only the former is suitable for readout. To
see this, notice that the detector measures the expectation value
of the sum of the island voltages in the $|0,1\rangle$ states:
$\langle \partial H/\partial n_+ \rangle = P_\phi^{ii} /M_\phi$,
$i=0,1$. However
\ba
 P^{00}_\phi = \sin \pi n_+ [A^{00}_{01} + A^{00}_{10} +
 (A^{00}_{23}+A^{00}_{32}) \sin^2 \pi
 n_- ], \nn
 P^{11}_\phi = \sin \pi n_+ [A^{11}_{01} + A^{11}_{10} +
 (A^{11}_{23}+A^{11}_{32}) \sin^2 \pi
 n_- ], \nonumber
\ea
both vanish when $n_+=0$ ($n_{gA}=-n_{gB}$), resulting in
indistinguishable (zero) outputs. Physically, the voltages of the
two island will be exactly equal but with opposite signs,
cancelling each other's effects on the output. Thus, the qubit
should be read out at $n_-=0$ (or any integer number). The
difference in the output between the two states is then
proportional to
\be
 P^{11}_\phi - P^{00}_\phi = \sin \pi n_+ [A^{11}_{01} + A^{11}_{10}
 - A^{00}_{01} - A^{00}_{10}], \label{P1100}
\ee
which vanishes at integer $n_+$, as expected.

Figure \ref{fig3} displays the result of numerical calculation of
$P_\phi^{11}-P_\phi^{00}$ for different values of $\alpha$, for
the same set of parameters as in Fig.~\ref{fig2}. The plot is vs
$n_+$ while $n_-{=}\ 0$ ($n_{gA}=n_{gB}$). The sinusoidal-type
behavior, is in agreement with (\ref{P1100}). Maximum output is
achieved when $n_+ {\approx}\ 0.5$ ($n_{gA}=n_{gB} \approx 0.25$).
This is indeed the optimal point for reading out the qubit.

None of the above arguments depended on the qubit's regime of
operation, although we mainly focussed on the flux regime. The
only restriction is that the qubit island voltages should not be
too small to be detectable by the SET; the qubit should not be too
far in the flux regime. It is indeed interesting to see how the
scheme works when the qubit is in charge regime ($E_C \gg E_J$).
Let us write the qubit wave function in charge basis
$|n_A,n_B\rangle$, where $n_{A,B}$ is the number of Cooper pairs
on the two islands.\cite{note2} When $n_{gA}=n_{gB}$ ($n_-=0$),
the states $|n,n+1\rangle$ and $|n+1,n\rangle$ will be degenerate.
The presence of the Josephson term of the third junction will
remove this degeneracy, making the rotated states $|n,\pm\rangle =
(|n,n+1\rangle \pm |n+1,n\rangle)/\sqrt{2}$ preferred basis for
the qubit. In the Hamiltonian matrix, $|n,+\rangle$ has nonzero
off-diagonal matrix elements with other charge states, hence will
mix after complete diagonalization. The matrix elements of
$|n,-\rangle$ with other states, however, are all zero as long as
$E_{J1}=E_{J2}$. It fact, $|n,-\rangle$ is an eigenstate of the
qubit Hamiltonian, as well as an eigenvector of $H_{\rm int}$ if
$\kappa_-=0$. Thus, $H_{\rm int}$ will not mix $|n,-\rangle$ with
any other states. It is easy to show that for all values of $n_+$,
one of the two lowest energy levels of the qubit will be a
$|n,-\rangle$ state (with an appropriate value of $n$ depending on
$n_+$), hence $\langle 1| H_{\rm int} |0\rangle = 0$.

To summarize, we have shown that QND measurement of a three
Josephson junction qubit is possible if both island charges are
measured symmetrically at a symmetric biasing point. The scheme is
applicable for a wide range of parameters from charge to flux
regime as long as charge measurement is possible. QND readout
allows long readout time with high fidelity and small
qubit-detector coupling to prevent decoherence due to the
interaction with the measurement device during quantum operation.
It also makes it possible to operate the SET in the quasiparticle
branch benefitting from larger charge sensitivity. Experimental
study of such a system can shine light on the intrinsic mechanisms
of relaxation and decoherence sources in charge or charge-flux
qubits.

The author is pleased to acknowledge discussion with A.J.~Berkley,
P.M.~Echternach, A.N.~Korotkov, A.~Maassen van den Brink,
A.Yu.~Smirnov, and A.M.~Zagoskin.


\begin{thebibliography}{99}


\bibitem{shnirman} Yu. Makhlin, G. Sch\"on, and A. Shnirman,
Phys. Rev. Lett. {\bf 85}, 4578 (2000); Rev. Mod. Phys. {\bf 73},
357 (2001).

\bibitem{note} For weak enough coupling, it is also possible to
continuously monitor the quantum evolution of the qubit. See e.g.
A.N. Korotkov and D.V. Averin, Phys. Rev. B {\bf 64}, 165310
(2001); A.Yu. Smirnov, Phys. Rev. B {\bf 68}, 134514 (2003); E.
Il'ichev {\it et al.},
Phys. Rev. Lett. {\bf 91}, 097906 (2003).

\bibitem{takayanagi} H. Tanaka, S. Saito, H. Nakano, K. Semba,
M. Ueda, H. Takayanagi, cond-mat/0407299.

\bibitem{QNDreview} V.B. Braginsky and F.Ya. Khalily,
Rev. Mod. Phys. {\bf 68}, 1 (1996).

\bibitem{QNDSchemes} See e.g. D.V. Averin, Phys. Rev. Lett.
{\bf 88}, 207901 (2002); A. Blais {\it et al.},
Phys. Rev. A {\bf 69}, 062320 (2004).

\bibitem{vion} D. Vion {\it et al.},
Science {\bf 296}, 886 (2002).

\bibitem{CPQ} M.H.S. Amin, Phys. Rev. B {\bf 71}, 024504 (2005).

\bibitem{mooij} J.E. Mooij {\it et al.},
Science \textbf{285}, 1036 (1999); T.P. Orlando {\it et al.},
Phys. Rev. B {\bf 60}, 15398 (1999).

\bibitem{rfSET} R.J. Schoelkopf {\it et al.},
Science {\bf 280}, 1238 (1998); M.H.
Devoret, and R.J. Schoelkopf, Nature {\bf 406}, 1039 (2000).


\bibitem{Cqubits} Such a readout scheme was proven
successful in charge qubits. See A. Guillaume {\it et al.},
Phys. Rev. B {\bf 69}, 132504 (2004); T. Duty {\it et al.},
Phys. Rev. B {\bf 69}, 140503 (2004).

\bibitem{orthodox} D.V. Averin and K.K. Likharov in
{\it Mesoscopic Phenomena in Solids}, edited by B. Altshuler {\it
et al.} (Elsevier, Amsterdam, 1991).

\bibitem{korotkov} A.N. Korotkov, Phys. Rev. B {\bf 49}, 10381 (1994)

\bibitem{DJQP} A.A. Clerk {\it et al.},
Phys. Rev. Lett. {\bf 89}, 176804 (2002).

\bibitem{note2} $H_Q$ and $H_{\rm int}$ are easily
obtained by replacing $P_\phi$ and $P_\theta$ with $n_A\pm n_B$,
respectively.


\end{thebibliography}
\end{document}